\documentclass{emulateapj}

\def\gsim { \lower .75ex \hbox{$\sim$} \llap{\raise .27ex \hbox{$>$}} }
\def\lsim { \lower .75ex \hbox{$\sim$} \llap{\raise .27ex \hbox{$<$}} }

\def\lesssim{\mathrel{\hbox{\rlap{\hbox{\lower4pt\hbox{$\sim$}}}\hbox{$<$}}}}
\def\gtrsim{\mathrel{\hbox{\rlap{\hbox{\lower4pt\hbox{$\sim$}}}\hbox{$>$}}}}
 
\shorttitle{The Stellar Populations of M31 Halo Substructure}
\shortauthors{Ferguson et al.}

\begin{document}
\slugcomment{ {\it Accepted for publication in the Astrophysical Journal Letters}}
\title{The Stellar Populations of M31 Halo Substructure\altaffilmark{1}}

\author{Annette M. N. Ferguson\altaffilmark{2,3},  Rachel A. Johnson\altaffilmark{4,5},
Daniel C. Faria\altaffilmark{4,6},  Mike J. Irwin\altaffilmark{7}, Rodrigo A. Ibata\altaffilmark{8}, Kathryn V. Johnston\altaffilmark{9}, Geraint F. Lewis\altaffilmark{10}, Nial R. Tanvir\altaffilmark{11}}
\altaffiltext{1}{Based on observations made with the NASA/ESA Hubble Space Telescope, obtained at the Space Telescope Science Institute, which is operated by the Association of Universities for Research in Astronomy, Inc., under NASA contract NAS 5-26555. These observations are associated with program GO9458.}
\altaffiltext{2}{Max-Planck-Institut f\"{u}r Astrophysik, Karl-Schwarzschild-Strasse 1, Garching bei M\"{u}nchen, D-85741  Germany}
\altaffiltext{3}{Present Address: Institute for Astronomy, University of Edinburgh, Blackford Hill, Edinburgh UK EH9 3HJ; ferguson@roe.ac.uk}
\altaffiltext{4}{European Southern Observatory, Alonso de Cordova 3107, Vitacura, Santiago, Chile}
\altaffiltext{5}{Department of Astrophysics, University of Oxford, Keble Road, Oxford UK OX1 3RH}
\altaffiltext{6}{Lund Observatory, Box 43, SE-221 00 Lund, Sweden}
\altaffiltext{7}{Institute of Astronomy, Madingley Road, Cambridge UK CB3 0HA}
\altaffiltext{8}{Observatoire de Strasbourg, 11, rue de l'Universit\'{e}, F-67000, Strasbourg, France}
\altaffiltext{9}{Astronomy Department, Wesleyan University, Middletown, CT 06459, USA}
\altaffiltext{10}{Institute of Astronomy, School of Physics, A29, University of Sydney, NSW 2006, Australia}
\altaffiltext{11}{Centre for Astrophysics Research, University of Hertfordshire, College Lane, Hatfield UK AL10 9AB}
\begin{abstract}
We present the first results from our survey of stellar substructure
in the outskirts of M31 using the {\sl Advanced Camera for Surveys}
(ACS) on board the Hubble Space Telescope.  We discuss the stellar
populations associated with five prominent stellar overdensities
discovered during the course of our panoramic ground-based imaging
survey with the Isaac Newton Telescope Wide-Field Camera (INT WFC); a
sixth pointing targets a region of $`$clean' halo.  The
colour-magnitude diagrams (CMDs), which contain between
$\approx$10,000--90,000 stars and reach four magnitudes below the
horizontal branch, reveal clear variations in morphology between most
fields, indicating that the age and/or metallicity mix of stars is not
constant at large radius.  This directly confirms the existence of
large-scale population inhomogeneities within the halo of M31 and
lends further support to the notion that M31 has formed, at least in
part, through satellite accretions.  We find a striking similarity
between the populations of the giant stellar stream and those of
another overdensity, the NE shelf, which lies north-east of the galaxy
center.  If these overdensities are associated with the same
population, then the difference in their red clump magnitudes implies
the NE shelf lies in front of the stream by several tens of kpc, in
good agreement with recent orbit calculations for the stream
progenitor.
\end{abstract}


\section{Introduction}
The INT WFC survey of M31 has led to the discovery of significant
stellar substructure in the halo and outer disk of our nearest giant
neighbour \citep{ibata01,ferg02,mcconn03,mcconn04,irwin05}.  At the
present time, the survey consists of more than $40$ square degrees of
V- and {\sl i}-band imaging to a depth sufficient to resolve stars to
$\approx 3$ magnitudes below the tip of the red giant branch (RGB).

Few constraints currently exist on the nature and origin of the
stellar substructure in M31. Within the popular $\Lambda$CDM model for
galaxy formation, substructure is expected in the outer regions of
galaxies as they continue to grow from the accretion and tidal
disruption of satellite companions. Recent calculations suggest that
massive disk galaxies, like the Milky Way and M31, could have
cannibalised $\sim100$ dwarf galaxy-like systems over their lifetimes
\citep{bull04}.  Since these accreted satellites are likely to have
experienced a range of star formation and chemical evolution
histories, one expects their tidal debris to be distinct in terms of
stellar populations.  If this picture is correct, then the spatial
substructure in M31 should be accompanied by stellar population
inhomogeneities.

In order to gain insight into the origin of the M31 halo substructure,
we are conducting a detailed study of the associated stellar
populations using the HST/ACS.  Our INT WFC survey indicated
intriguing variations in the mean RGB colour across the halo, which
could reflect changes in metallicity and/or age.  HST is required to
confirm these variations and, via detection of fainter and more
evolved populations such as the red clump and horizontal branch,
provide constraints on their nature.  In this {\it Letter}, we present
the first results from our HST/ACS survey which establish clear
differences, as well as some similarities, between the populations in
various substructures.

\section{Observations}

Observations at six locations in the outskirts of M31 were obtained
during Cycle 11 with the Wide Field Camera of the ACS as part of
GO\#9458.  Our fields sample the five most prominent stellar
overdensities in the maps of \cite{ferg02} as well as a region of
apparently $`$clean' halo (\texttt{Minor Axis}) situated 20~kpc along
the southern minor axis (see Figure 1).  The main characteristics of
the stellar overdensities can be summarized as follows:

{\noindent\texttt{Giant Stream} -}~~a stellar stream, with a
well-defined eastern edge, which can be traced in projection to $\sim
70$~kpc from the center of the galaxy towards the south-east
\citep{ibata01,mcconn03}. Radial velocity data have ruled out a simple
connection to the dwarf elliptical companions M32 and NGC~205
\citep{ibata04,font04};\\
{\noindent\texttt{NE Shelf} -}~~a large
overdensity north-east of the galaxy centre which also exhibits a sharp outer boundary;\\
{\noindent\texttt{NGC~205 Loop} -}~~a small ($\sim15$~kpc) loop of
stars which appears, in projection, to emanate from NGC~205. While the
geometry is very suggestive of a connection between the two, extant
stellar kinematical data still leave open other
possibilities \citep{mcconn04}; \\
{\noindent\texttt{G1 Clump} -}~~a stellar overdensity which sits
30~kpc along the south-western major axis, in close proximity to the
anomolous globular cluster G1.  Although initially a connection
between the two was suspected, radial velocity
data have now ruled this out \citep{reitzel04,ferg04};\\
{\noindent\texttt{Northern Spur} -}~~first hinted at by \cite{wk88},
this is an extended plume of stars which sits just above the
north-eastern major axis.

\begin{figure}
\includegraphics*[angle=0,scale=0.50]{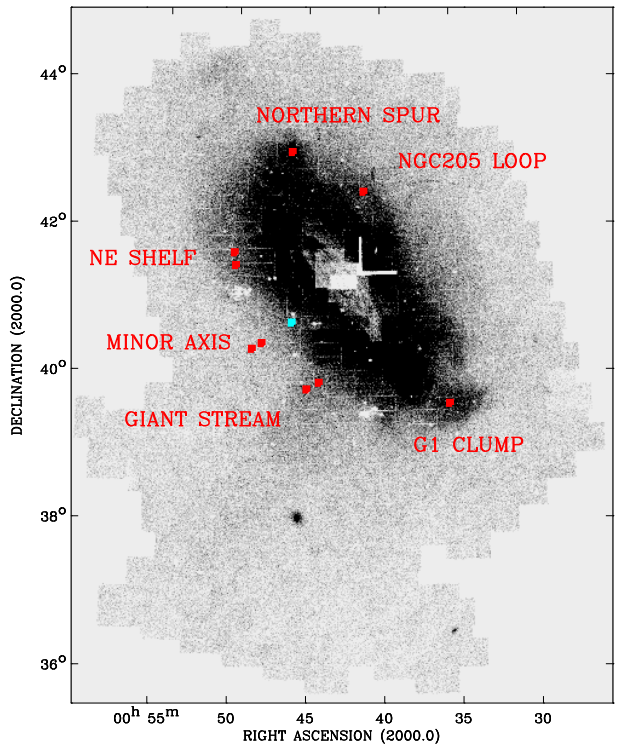}
\caption{A map of the surface density of blue RGB stars in a
$40\sq\arcdeg$ area around M31 as mapped with the INT WFC. Overlaid
are the locations of the HST/ACS pointings presented here (red) and the 
deep HST/ACS halo pointing of \cite{brown03} (cyan).}
\end{figure}

 Each field was observed for one orbit in
the F606W filter (broad V) and two orbits in the F814W filter
(I), all within a single visit. Integer pixel dithers were executed between multiple
sub-exposures to facilitate warm pixel and cosmic-ray rejection.
Given the low stellar density in the outer halo, two separate
pointings were obtained in each of the three outermost fields
(\texttt{Minor Axis}, \texttt{NE Shelf} and \texttt{Giant Stream}) in
order to ensure sufficient statistics on the upper RGB and horizontal
branch.

Images were first processed through the ACS pipeline and those in a
given passband were combined using the Multidrizzle task
\citep{koek02} within PyRAF. We used the default mode of Multidrizzle
which calculates the offsets between dithered exposures using the
information in the image header world coordinate system.  The final
drizzle was conducted using the Lanczos3 kernel with {\it pixfrac} and
{\it scale} set to unity.

Photometry was subsequently obtained using the IRAF implementation of
DAOPHOT \citep{stet87}.  The stellar density in our fields is low
enough to allow precise results from aperture photometry alone.
PSF-fitting photometry was also carried out to compare with the
aperture photometry and to measure how accurately individual sources could be
modelled by a stellar PSF.  For this purpose, a spatially non-varying
PSF was built using $\sim150$ bright stars in each of the nine ACS
pointings.  Sources were retained in our final photometry list if
their magnitude errors, sharpnesses and chi values all lay within
3-sigma of the average value at their magnitude.  Aperture corrections
were derived from several tens of bright stars in each combined frame
and our photometry was placed on the STMAG system using the latest
zeropoint values \citep{siri05}.  Inspection of the \cite{schlegel98}
reddening map indicates values of E(B$-$V)$=0.06-0.08$ (with
uncertainty 16\%) in the far outskirts of M31; we adopt a constant
value of E(B$-$V)$=0.07$ for all fields and values of the selective
extinction coefficient in the F606W and F814W passbands from
\cite{siri05}.  The final CMDs contain between 10,000 stars (the
\texttt{Minor Axis} field) and 90,000 stars (the \texttt{Northern
  Spur} field) and extend to four magnitudes below the horizontal
branch (see Figure 2).  Artificial star tests indicate 75\%
completeness to F814W$_0=28.5$ in fields of average stellar density.
Details of our analysis will be presented in a forthcoming paper.

\section{Results}

\subsection{The CMD Morphologies}

\begin{figure*}
\includegraphics*[angle=90,scale=0.7]{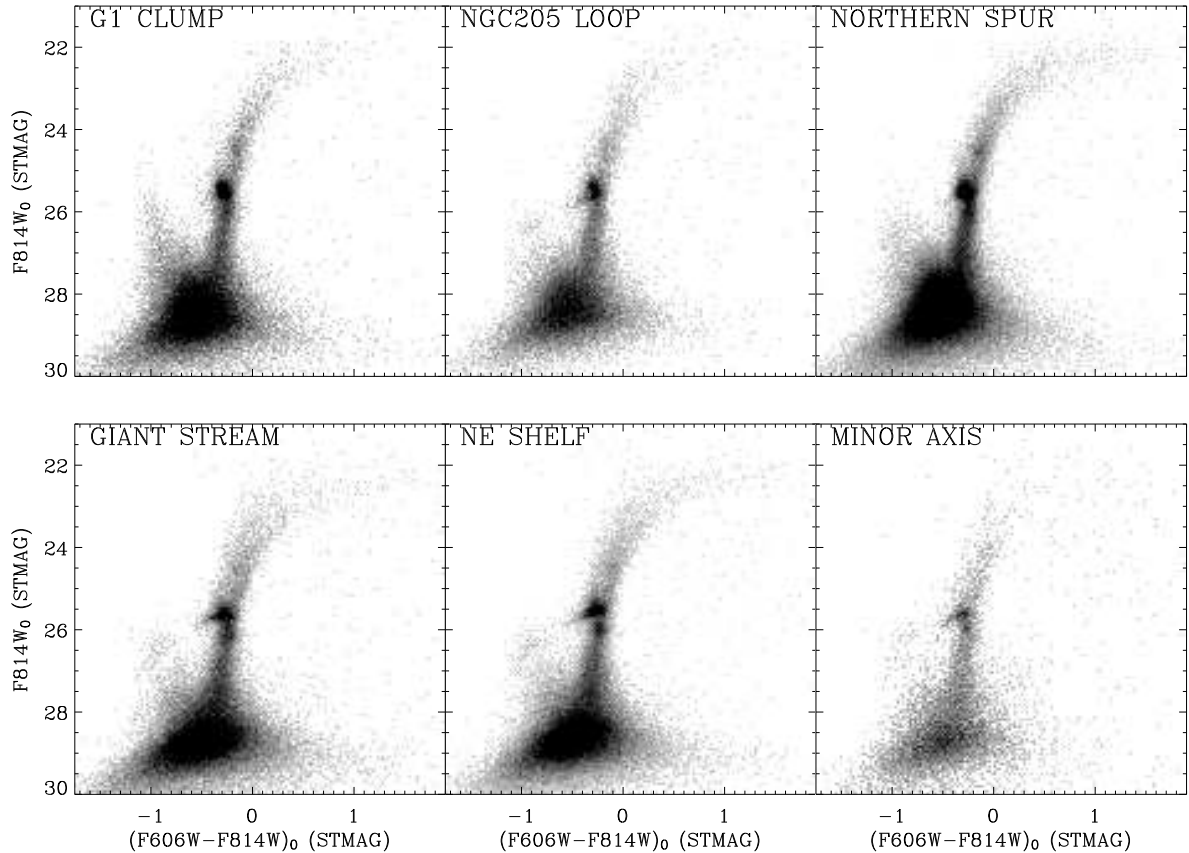}
\caption{Hess diagrams of six halo fields in M31 shown with a square
root stretch. \texttt{G1 Clump, NGC205 Loop, Northern Spur, Giant
Stream} and \texttt{NE Shelf} are regions of stellar substructure within
the M31 halo, while \texttt{Minor Axis} was chosen to represent clean
halo.  Clear differences are apparent between most CMDs, indicating
genuine stellar population variations within the halo substructure.}
\end{figure*}

Figure 2 shows Hess diagrams for the six halo fields targetted in our
survey.  In all cases, a broad RGB and a prominent red clump are
evident, consistent with the findings of earlier HST studies of the
M31 halo (e.g. \cite{holland96,bella03}). However, on closer
inspection, noticeable differences can be seen in the morphology of
many of the CMDs.  For example, the \texttt{Giant Stream} and
\texttt{NE Shelf} exhibit rather prominent RGB bumps (F814W$_0 \approx
26.1$). Note that the lower stellar density of the \texttt{Minor Axis}
CMD makes it difficult to assess whether the RGB bump is similarily
strong in this field.  The RGB bump arises due to an evolutionary
pause on the first ascent RGB when the H-burning shell passes through
the discontinuity left by the deepest penetration of the convective
envelope \citep{thomas67,iben68}.  It is a short-lived phase and,
although commonly seen in single-age populations, has only been
recently identified in composite populations (e.g. \cite{monaco02}).
The clarity of this feature in the shelf and stream CMDs suggests
these regions contain a dominant stellar population with a small to
moderate range of age and metallicity, perhaps the result of one or
more bursts.  Additionally, the location of the RGB bump at
approximately 0.4 magnitudes fainter than the red clump in these
fields supports the inference from the RGB colour that this is a
relatively metal-rich population (e.g. \cite{alves99}).

The \texttt{G1 Clump} is distinct in showing clear evidence for a
young population.  It exhibits a well-populated upper main sequence
(blue plume), with some stars as bright as F814W$_0 \approx 24.5$.
Comparison to the Padova isochrones \citep{gir04} indicates the
brightest stars are consistent with a population of age of
$\sim2.5\times10^8$ years \citep{faria05}.  On the other side of the
major axis,and at a slightly smaller radius, the \texttt{Northern
  Spur} shows no evidence for such a young component.  Instead, this
field shows a low-amplitude RGB bump at F814W$_0\sim26$ and another
bump at $\sim24.5$.  Given the moderate metallicity inferred from the
RGB colour, the models of \cite{alves99} indicate that the brighter
bump is more likely to be an asymptotic giant branch (AGB) bump as
opposed to a secondary RGB bump (such features are also marginally
present in the \texttt{G1 Clump} CMD).  The \texttt{NGC~205 Loop} is
similar to the \texttt{Giant Stream}, \texttt{NE Shelf} and
\texttt{Minor Axis} fields in having an extended blue horizontal
branch, but it differs in having neither an RGB/AGB bump nor a blue
plume. Finally, all three CMDs in the top panel of Figure 2 show
tantalizing evidence of a bright main-sequence turn off at F814W$_0
\approx 27.5$, roughly matched by a population of $\sim2.5\times10^9$
years with [Fe/H]$=-0.7$ \citep{gir04}.  As the \texttt{Giant Stream} and
\texttt{NE Shelf} fields contain more stars than either the \texttt{G1 Clump}
or \texttt{NGC~205 Loop}, the visibility of this putative turnoff is not
simply dependent on total stellar density.

Although the CMD morphologies often differ, the mean colour of
the (broad) RGB remains surprisingly constant.  When measured in the
region $27<\rm{F814W_0}<26$, the mean RGB colour varies
by less than 0.05 magnitudes from region to region.  This is
comparable to the difference expected in colour between [Fe/H]$=-0.4$ and $-0.7$ 
stellar isochrones for a population of 14~Gyr \citep{gir04}.

\subsection{Using Stellar Populations to Trace the Giant Stream}

The striking resemblance between the \texttt{Giant Stream} and
\texttt{NE Shelf} CMDs motivates further exploration of the
relationship between these features.  Figure 3 shows the F814W$_0$
luminosity function (LF) of these populations in the vicinity of the
red clump and RGB bump.  We fit the LF in the red clump region with
the sum of a first order polynomial and a gaussian, selecting stars
with the same range of de-reddened magnitude and colour in each case.
We find the peak magnitude of the red clump to differ slightly between
the fields, being F814W$_0=25.67\pm0.01$ for the stream and $25.53\pm0.01$
for the shelf.

The magnitude and colour of the red clump are known to vary as a
function of age and metallicity (e.g. \cite{gir01}) but the overall
similarity of the CMD morphologies suggests that population-driven
differences between the stream and shelf regions are small.  This is
further confirmed by the nearly identical luminosity function slopes
observed.  In this case, a more likely cause of the offset in red
clump magnitude is that the regions are located at different
line-of-sight distances.  The red clump magnitude difference of
$\Delta({F814W_0})_{STREAM-SHELF}=0.14\pm0.01$ implies that the
\texttt{Giant Stream} region lies beyond the \texttt{NE Shelf} by a
factor of $1.07\pm0.01$.  \cite{mcconn03} have measured the
line-of-sight distance to the stream as a function of position along
it.  We have interpolated their measurements to the location of our
\texttt{Giant Stream} ACS field and find a line-of-sight distance of
$830\pm20$~kpc.  Combining this value with the differential distance
modulus inferred from the red clump magnitudes yields a distance to
the shelf of $776\pm20$~kpc.

\section{Discussion}

Our HST/ACS survey of M31 substructure has revealed significant
variations in CMD morphology at large radius, indicating that the age
and/or metallicity of stars in these parts is not constant. This is the
first direct evidence for large-scale population inhomogeneities in
the outskirts  of M31. Prior HST studies, which have inferred a 
uniform population,  have generally been based on random
pointings within the inner halo and have missed the main stellar
overdensities (e.g. \cite{bella03}).

It is tempting to interpret the observed population variations as
evidence for multiple satellite accretions within the M31 halo, each
having a distinct star formation and chemical evolution history.  One
puzzle, however, is the rather uniform mean RGB colour (though broad
distribution) observed from field-to-field.  If interpreted purely in
terms of metallicity, the mean RGB colour suggests a variation of
$\lesssim0.3$~dex throughout the substructure.  This could suggest
that much of the halo debris comes from the accretion of a single
system -- one which has undergone little chemical evolution during
several orbits around M31 but experienced at least one burst of star
formation.  On the other hand, age and metallicity may be varying in
tandem throughout these regions, conspiring to produce a
remarkably constant mean RGB colour (for example, as seen in the
Carina dwarf spheroidal galaxy (e.g. \cite{rizzi03})).  We will evaluate
these scenarios in more detail in subsequent papers where we will
model the CMDs in terms of star formation and chemical evolution
histories.

\begin{figure}
\includegraphics*[angle=90,scale=0.40]{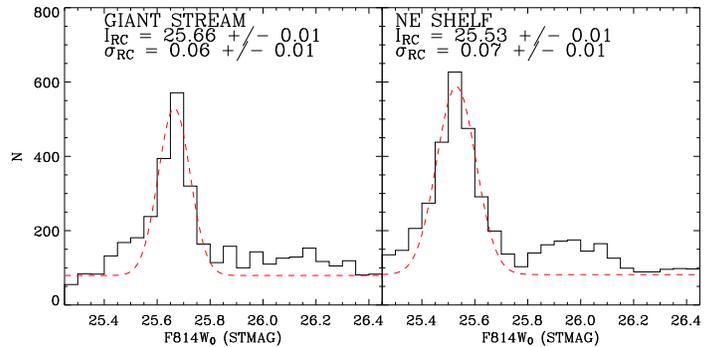}
\caption{F814W$_0$ LFs in the vicinity of the red clump and RGB bump
for the \texttt{Giant Stream} (left) and \texttt{NE Shelf}
(right). Overlaid (dashed line) are gaussian fits to the red clump. }
\end{figure}

Previous studies of the M31 inner halo have demonstrated that
$\sim$25-50\% of the population is old and metal poor (e.g.
\cite{holland96,brown03}).  The presence of a blue horizontal branch
in four of the six CMDs in Figure 2 suggests this smooth halo population is also
present at larger radii.  The \texttt{Northern Spur}\ and \texttt{G1
  Clump} CMDs, which probe the most extreme radii in our sample, do
not, however, show this feature.  Assuming our \texttt{Minor Axis}\ 
field is representative of the smooth halo component in M31, we can
ask whether we should see this component in the outermost
fields studied.  We find that, for both a spherical and flattened
(b/a$=0.6$) R$^{1/4}$ distribution, the amount of smooth halo expected
in the \texttt{Northern Spur}\ and \texttt{G1 Clump} CMDs is almost
entirely negligible compared to the number of stars in the overdensity
(4--10\% of the detected stars). In other words, the absence of an obvious extended blue
horizontal branch in these fields does not necessarily mean that the smooth 
halo component seen at smaller radii  is not also present.

The currently-favoured orbit for the stream progenitor is highly
radial and passes very close to the galaxy center before re-emerging
north-east of the nucleus \citep{ibata04,font04}. Although radial
velocities have not yet been published for stars in the \texttt{NE
  Shelf} region, the strong similarity between the stream and shelf
CMDs found here appears to support this prediction.
Furthermore, the calculations of \cite{ibata04} and \cite{font04}
indicate that the shelf region probed by our ACS pointing should lie
many tens of kpc in front of our stream pointing.  Our differential
distance estimate based on the red clump magnitude is $\approx
50\pm30$~kpc and is thus in encouraging agreement.

\cite{brown03} have recently obtained a very deep CMD of the M31 halo
at a location 11~kpc along the southern minor axis (see Figure 1). The
data reveal the surprising discovery of a significant (30\% by mass)
intermediate-age (6--8~Gyr) metal-rich population -- a result somewhat
at odds with the traditional picture of stellar halos as old and
metal-poor.  The results presented in this {\it Letter} suggest
caution in interpreting this finding, as well as all others that are
based on observations of small single fields. The stellar populations
in the outskirts of M31 exhibit large-scale variations and hence
inferences about halo formation drawn from such pointings may not be
representative of the entire halo.  This is especially true in the
case of Brown's field, which lies in close proximity to two regions
which are strongly contaminated by stream debris.

\acknowledgements
 Leo Girardi, Marco Sirianni and Tom Brown are thanked for providing
 results ahead of publication.  AMNF acknowledges support from a Marie
 Curie Fellowship of the European Community under contract number
 HPMF-CT-2002-01758 and the ESO Santiago visitor program.  Support for
 program GO9458 was provided by NASA through a grant from the Space
 Telescope Science Institute, which is operated by the Association of
 Universities for Research in Astronomy, Inc., under NASA contract NAS
 5-26555.

\end{document}